\documentclass[aps,prd,twocolumn,superscriptaddress,showpacs,amsmath,amssymb]{revtex4}

\usepackage{graphicx,amssymb}
\usepackage{color}
\usepackage{comment}
\usepackage{multirow}
\usepackage{dcolumn}
\usepackage{bm}
\includecomment{printrobustnotes}
\includecomment{printallnotes}
\usepackage{xspace}

\def\rntt{$^{222}$Rn~}

\def\pbten{$^{210}$Pb~}
\def\bitwoonefour{$^{214}$Bi~}
\def\pbsix{$^{206}$Pb~}
\def\poten{$^{210}$Po~}
\def\rntte{$^{222}$Rn}

\def\pbtene{$^{210}$Pb}
\def\pbsixe{$^{206}$Pb}
\def\potene{$^{210}$Po}

\usepackage{romanbar}
\usepackage{graphicx}
\usepackage{dcolumn}
\usepackage{bm}

\usepackage{amstext}

\begin{document}

\title{Measurement of low-energy events due to \rntt daughter contamination on the surface of a NaI(Tl) crystal}

\author{K.W. Kim}
\affiliation{Center for Underground Physics, Institute for Basic Science (IBS), Daejeon 34126, Korea}
\author{C. Ha}
\affiliation{Center for Underground Physics, Institute for Basic Science (IBS), Daejeon 34126, Korea}
\author{N.Y. Kim}
\affiliation{Center for Underground Physics, Institute for Basic Science (IBS), Daejeon 34126, Korea}
\author{Y.D. Kim}
\affiliation{Center for Underground Physics, Institute for Basic Science (IBS), Daejeon 34126, Korea}
\affiliation{Department of Physics and Astronomy, Sejong University, Seoul 05006 , Korea}
\affiliation{University of Science and Technology (UST), Daejeon 34113, Korea}
\author{H.S. Lee}
\email{hyunsulee@ibs.re.kr}
\affiliation{Center for Underground Physics, Institute for Basic Science (IBS), Daejeon 34126, Korea}
\affiliation{University of Science and Technology (UST), Daejeon 34113, Korea}
\author{B.J. Park}
\affiliation{Center for Underground Physics, Institute for Basic Science (IBS), Daejeon 34126, Korea}
\affiliation{University of Science and Technology (UST), Daejeon 34113, Korea}
\author{H.K. Park}
\affiliation{Department of Accelerator Science, Korea University, Sejong 30019, Korea}

\begin{abstract}
		It has been known that decays of daughter elements of \rntt on the surface of a detector cause significant background at energies below 10~keV.
In particular \pbten and \poten decays on the crystal surface result in significant background for dark matter search experiments with NaI(Tl) crystals. 
In this report, measurement of \pbten and \poten decays on surfaces are obtained by using a \rntt contaminated crystal. 
Alpha decay events of \poten on the surface are measured by coincidence requirements of two attached crystals. Due to recoiling of \pbsixe, rapid nuclear recoil events are observed. 
A mean time characterization demonstrates that \pbsix recoil events can be statistically separated from those of sodium or iodine nuclear recoil events, as well as electron recoil events. 

\end{abstract}

\maketitle

\section{Introduction}
Numerous astronomical observations suggest that nonbaryonic cold dark matter is the dominant form of matter in the Universe~\cite{Komatsu:2010fb,Ade:2013zuv}. The weakly interacting massive particle~(WIMP) is one of the most prominent candidates supported by astronomical observations and particle physics theories that extend the standard model~\cite{lee77,jungman96}. Various experimental studies on WIMP dark matter have been conducted by detecting signals from recoiling nuclei ~\cite{gaitskell04,baudis12}. Among these experiments, a DAMA/LIBRA experiment is particularly interesting due to the positive signals observed for an annual modulation of event rates that were observed with an array of NaI(Tl) crystals~\cite{dama}. This modulation signal has been a subject of continuing debate because other experiments have observed null signals in the regions of the WIMP-nucleon cross-section and WIMP-mass parameter space favored by DAMA/LIBRA observations ~\cite{kims_csi,supercdms,lux,xenon_am,xenon-1t}. However, there is room to explain all of the direct search results in terms of non-trivial systematic differences in detector responses and possible modifications of the commonly used halo model for the galactic distribution of dark matter~\cite{sys1,sys2}. It is, therefore, required that an experiment similar to the DAMA/LIBRA should use the same NaI(Tl) target detectors. To verify the observations from the DAMA/LIBRA experiments, a few experimental efforts regarding the NaI(Tl) detector have been carried out recently~\cite{kims_nai,dm-ice,anais,sabre,cosine}. 

An annual modulation analysis from the other NaI(Tl) experiments will allow a direct comparison of the modulation amplitude. This minimizes systematic differences between the two experiments and enables a model-independent comparison. 
However, it is also interesting to extract nuclear recoil events using the NaI(Tl) detectors, regardless of whether the DAMA/LIBRA signals originated from WIMP-nucleon interaction. Taking advantage of the high light output from the recently developed NaI(Tl) crystals~\cite{kims_nai,anais_pe}, a good pulse shape discrimination~(PSD), where the nuclear recoil events have faster decay time than the electron recoil events, has been achieved with a NaI(Tl) crystal detector~\cite{kims_psd}. 

A dangerous background event is observed while searching for WIMP dark matter and is produced due to the \rntt progenies on the detector surface, as reported in various experiments~\cite{cdms_pb210,cresst_pb210,kims_sa}. 
Among the progenies of \rntte, \pbten and \poten are the most problematic radioisotopes because of their long half-lives of 22.3 years and 138 days, respectively.
$\alpha$-decay of the \poten on the crystal surface also generates anomalously fast events~\cite{kims_sa,naiad_sa} due to heavy nuclei recoil from the interaction with \pbsixe. 
If an $\alpha$ particle escapes the crystal surface without energy deposition, the recoiled \pbsix with 103~keV kinetic energy creates a recoil signature on the crystal surface, which can mimic the nuclear recoil events and cause them to be misidentified as a WIMP-nucleon interaction. It is then required to understand the \pbsix surface recoil for a direct extraction of the WIMP-nucleon interaction events using the PSD analysis~\cite{kims_sa}.

It is interesting to explore the distribution of $\beta$-decay from surface \pbten because it is reported as one of the dominant backgrounds in the low energy region for NaI(Tl) crystals~\cite{kims_jeon, anais_bg}. 
The modeling of the \pbten surface spectrum is limited by a lack of knowledge about the depth distribution of \pbten on the crystal surface. 
Data obtained with a \pbten surface contaminated crystal could help us to understand the depth profile of the surface of \pbtene. 
In this article, we present measurements of mean time characteristics as well as energy spectra from the \pbsix surface recoil events and the surface \pbten $\beta$-decay events using a \rntt contaminated crystal. 

\section{Experimental Setup}

\begin{figure}[!htb]
\begin{center}
		\includegraphics[width=8cm]{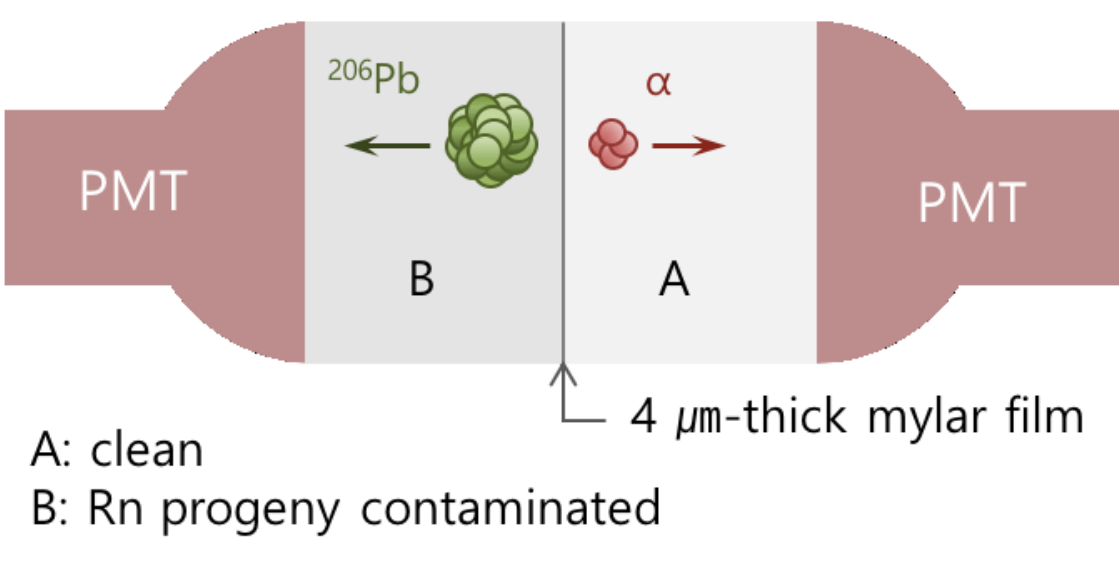}
		\caption{Schematic design of the detector. The \rntt contaminated crystal~(Crystal B) and clean crystal~(Crystal A) were directly attached with 4$\mu$m aluminum mylar film.}
\label{detector} 
\end{center}
\end{figure}

A cylindrical NaI(Tl) crystal of 8~cm diameter, 10~cm length, and a total mass of 1.28~kg was grown and encapsulated by Alpha Spectra Inc. and was used for the experiments. 
This is the same ingot crystal of two full-size NaI(Tl)~(C6 and C7) used for the COSINE-100 experiment~\cite{cosine}.
The crystal was cut into two equal pieces. One piece~(Crystal A) was stored in a clean environment after cleaning and polishing the surface. The other piece~(Crystal B) was irradiated 
with \rntt progenies by placing it in a \rntt contaminated chamber for two weeks~(April 15 -- April 30, 2017). The \rntt concentration was initially around 5~MBq/m$^{3}$.  
The two crystals were attached face to face at the cut surface as one can see in Fig.~\ref{detector}. A 4~$\mu$m aluminum mylar film was inserted between the two crystals to avoid cross-talk of scintillating light.

A 3-inch Hamamatsu R12669 photomultiplier tube~(PMT) was attached to each crystal for detecting scintillating photon signals. 
The signal from each PMT was amplified by home-made preamplifiers. The high-gain anode signal for low energy events was amplified by a factor of 30, and the low-gain 5$^{th}$ stage dynode signal for high energy events was amplified by a factor of 100. 
The amplified signals were digitized by a 500~MHz, 12-bit flash analog-to-digital converters~(FADCs). 
We applied two different triggers for the \pbsix recoil events and surface \pbten events in which each measurement with different trigger was done independently.
For the \pbsix recoil events, it was required that the single photoelectrons from the two crystals be coincident within a 200~ns window in the anode channels. 
Both the low-gain dynode and high-gain anode waveforms were recorded if the trigger conditions were accepted. 
Because the \pbten $\beta$-decay deposits usually hit only a single crystal, more than three photoelectrons were required to occur within a 200~ns time window for single channel.
If one crystal accepted this condition, then all crystals accepted data. 
The energy scales for both crystals were calibrated using 59.54 keV $\gamma$ events from a $^{241}$Am source. Internal 609 keV $\gamma$ events from \bitwoonefour decay were used for high energy calibration.

\section{Data Analysis}
Data was collected for a period of three months. Collection using the coincident trigger condition began approximately one month after the \rntt contamination for the \pbsix recoil events. 
In the series of \rntt decays, the longest half-life element before \pbten is \rntt with 3.8~days. About one month later, the only available \rntt progenies are \pbten and its daughter radioisotopes. 
\pbten decays into \poten as a $\beta$-decay with a 22.3 year half-life. Due to the 138~days half-life of \potene, one can expect an increased rate of \poten decays with a time constant of its life-time. 
\poten undergoes $\alpha$ decays into \pbsix and an $\alpha$ particle with a Q-value of 5.407~MeV. The kinetic energy of \pbsix and the $\alpha$ particle are 103~keV and 5.304~MeV, respectively. 
Because both \pbsix and $\alpha$ are heavy nuclei, these signals create signatures similar to the nuclei's recoil events. If an energetic $\alpha$ particle escapes the crystal surface without energy loss,
the recoiling \pbsix creates low energy nuclear recoil signatures. These signatures can mimic the WIMP-nuclei's interactions.

\begin{figure}[!htb]
\begin{center}
		\includegraphics[width=8cm]{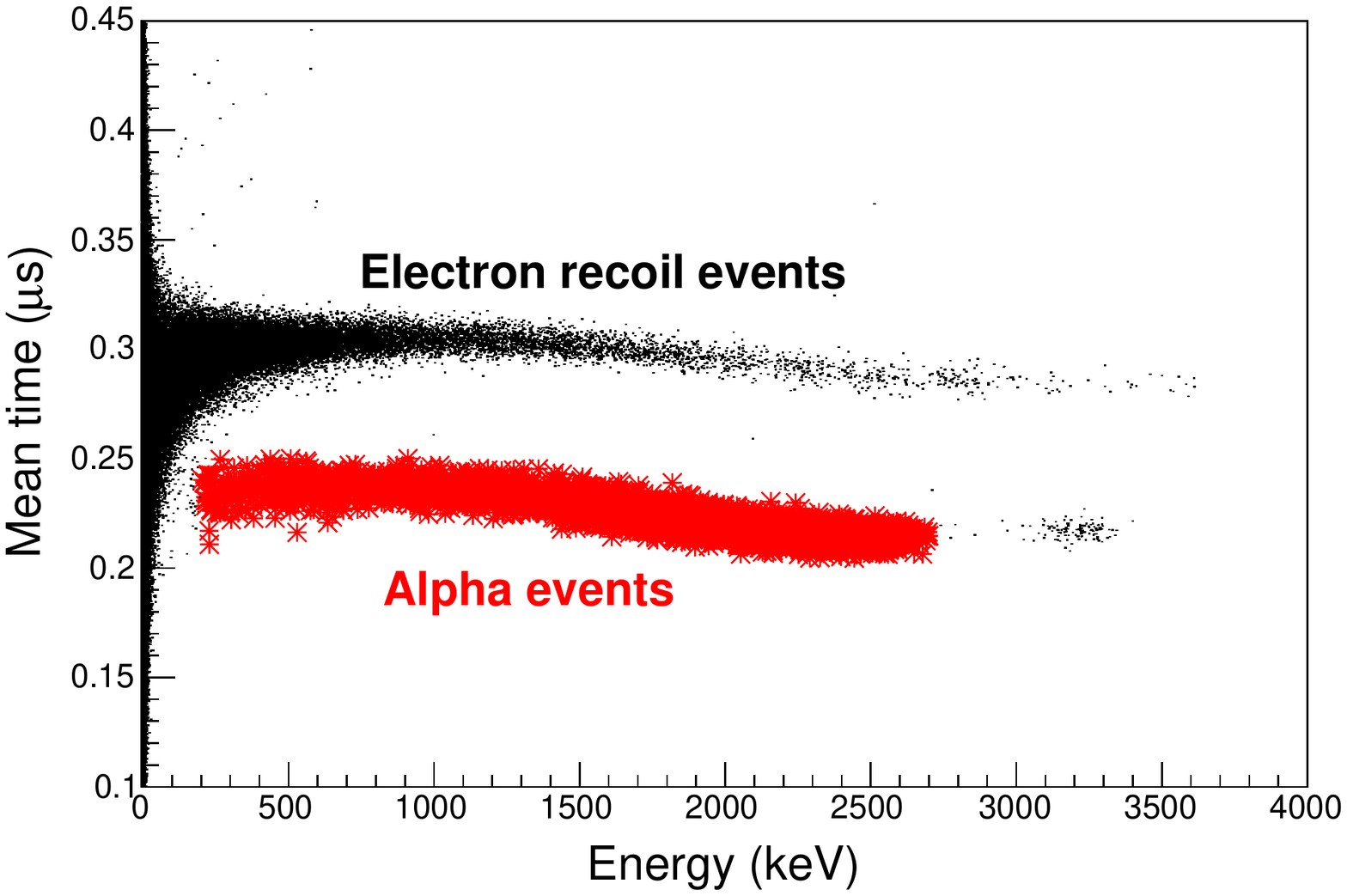}
		\caption{Surface $\alpha$ tagging with the clean crystal A using mean time distribution. The energy is given in keV electron equivalent energy where true energy of $\alpha$ are reduced with 50--70\% of $\alpha$/$\beta$ light ratio in NaI(Tl) crystals~\cite{naialpha}. }
\label{pb206_nEMT} 
\end{center}
\end{figure}

\begin{figure}[!htb]
\begin{center}
		\includegraphics[width=8cm]{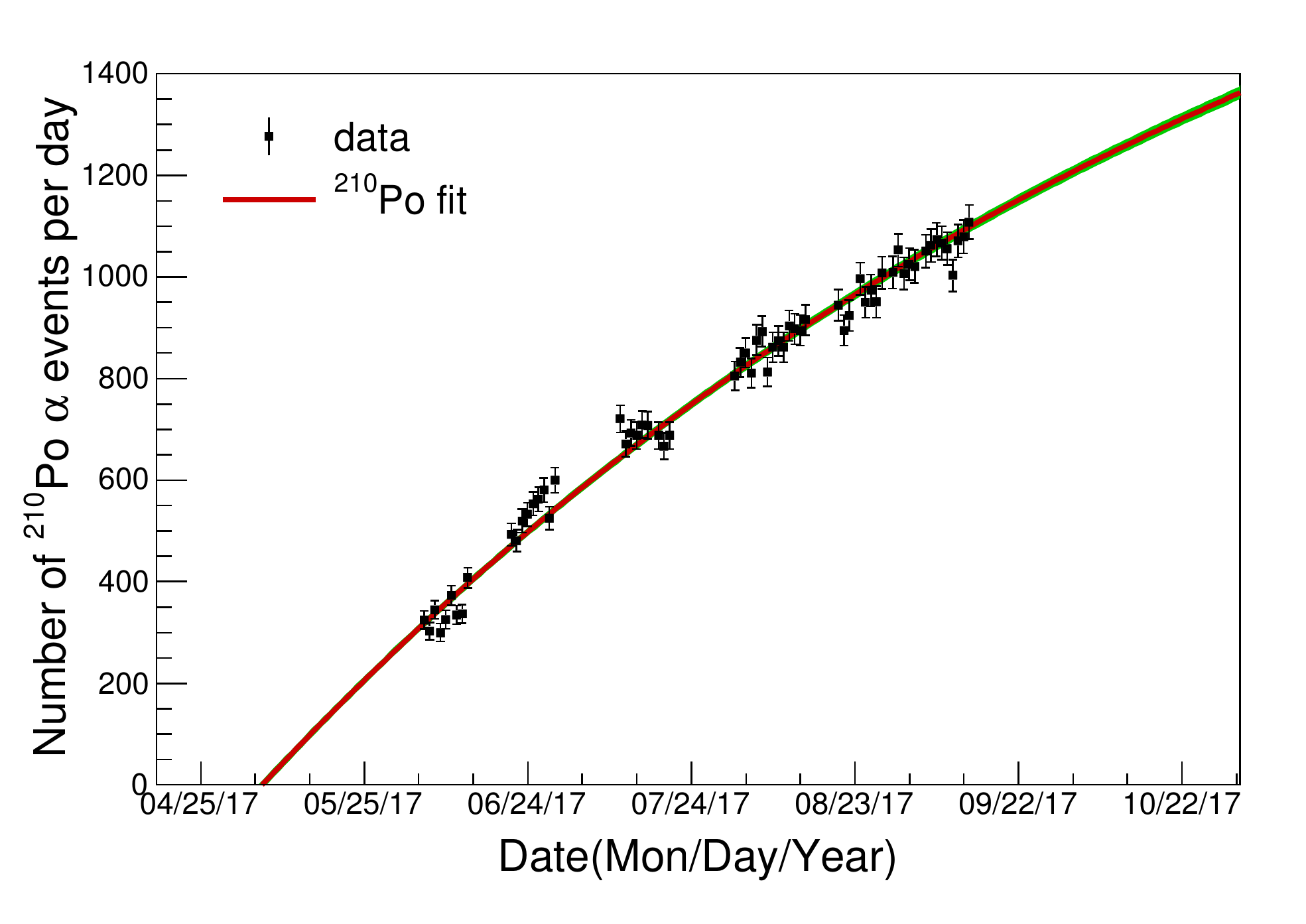}
		\caption{ Number of selected surface $\alpha$ events from the clean crystal A as a function of time. The black dots represent the observed surface $\alpha$ rate. This data is fit to an exponential function with a half-life of 138 days in \poten and is shown using a red line. 
}
\label{alpharate}
\end{center}
\end{figure}

To characterize \pbsix recoil events in the contaminated crystal B, we tagged an escaped $\alpha$ particle using the clean crystal A. One can identify the surface $\alpha$ events by taking advantage of the good pulse shape discrimination~(PSD) between electron recoil events and $\alpha$ recoil events in the NaI(Tl) crystal, as shown in Fig.~\ref{pb206_nEMT}. 
Here we used the mean time as a PSD parameter, which is defined as 
\begin{equation}
\mathrm{Mean~Time}=\left(\frac{\sum{A_it_i}}{\sum{A_i}}-t_0\right), 
\end{equation}
where $A_i$ and $t_i$ is the charge and time of the $i$th cluster~(for low energy anode signals) or digitized bin~(for high energy dynode signals), and $t_0$ is the time of the first cluster or first bin (above the threshold). Each cluster 
is identified by applying a clustering algorithm to the raw spectrum of the data~\cite{kims_1st}. 
Due to the thin mylar layer as well as small energy depositions on the contaminated crystal B, the measured energy of the $\alpha$ particles
is below the full energy deposition of the \potene. We exclude $\alpha$ events around 3200~keV electron equivalent energy region because true energy is approximately full energy deposition of \poten (5407~keV) considering 60\% of $\alpha$/$\beta$ light ratio in this energy~\cite{naialpha}. The $\alpha$ events with full energy deposition were due to internal \pbten contamination in the crystal~\cite{kims_nai}. The rate of measured $\alpha$ particles below full energy in crystal A, corresponding to the escaped $\alpha$ particles from the crystal B, has been monitored as a function of time and is shown in Fig.~\ref{alpharate}. In this figure, it is evident that the rate of surface $\alpha$ events has increased with the life-time of \potene. This can be modeled with the following formula,
\begin{equation}
		R_{\alpha}(t) = A (1-e^{-(t-t_0)/\tau_{Po_{210}}}), 
\end{equation}
where $t_0$ is the time when the initial \pbten contamination occurred. In the fit, the \poten half-life, $\tau_{Po_{210}}$=138~days, is fixed. Obtained $t_0$, May 5th with 2~days uncertainty, is reasonably close to the exposure period of \rntt contamination. The discrepancy between the best-fit value and the known exposure time is partially explained by an accumulation time of \pbten with the longest half-life of 3.8 days in \rntte. Therefore, the $\alpha$ particles with partial energy deposition in the clean crystal A, correspond to \poten decays on the surface of the contaminated crystal B.

\begin{figure}[!htb]
\begin{center}
		\includegraphics[width=8cm]{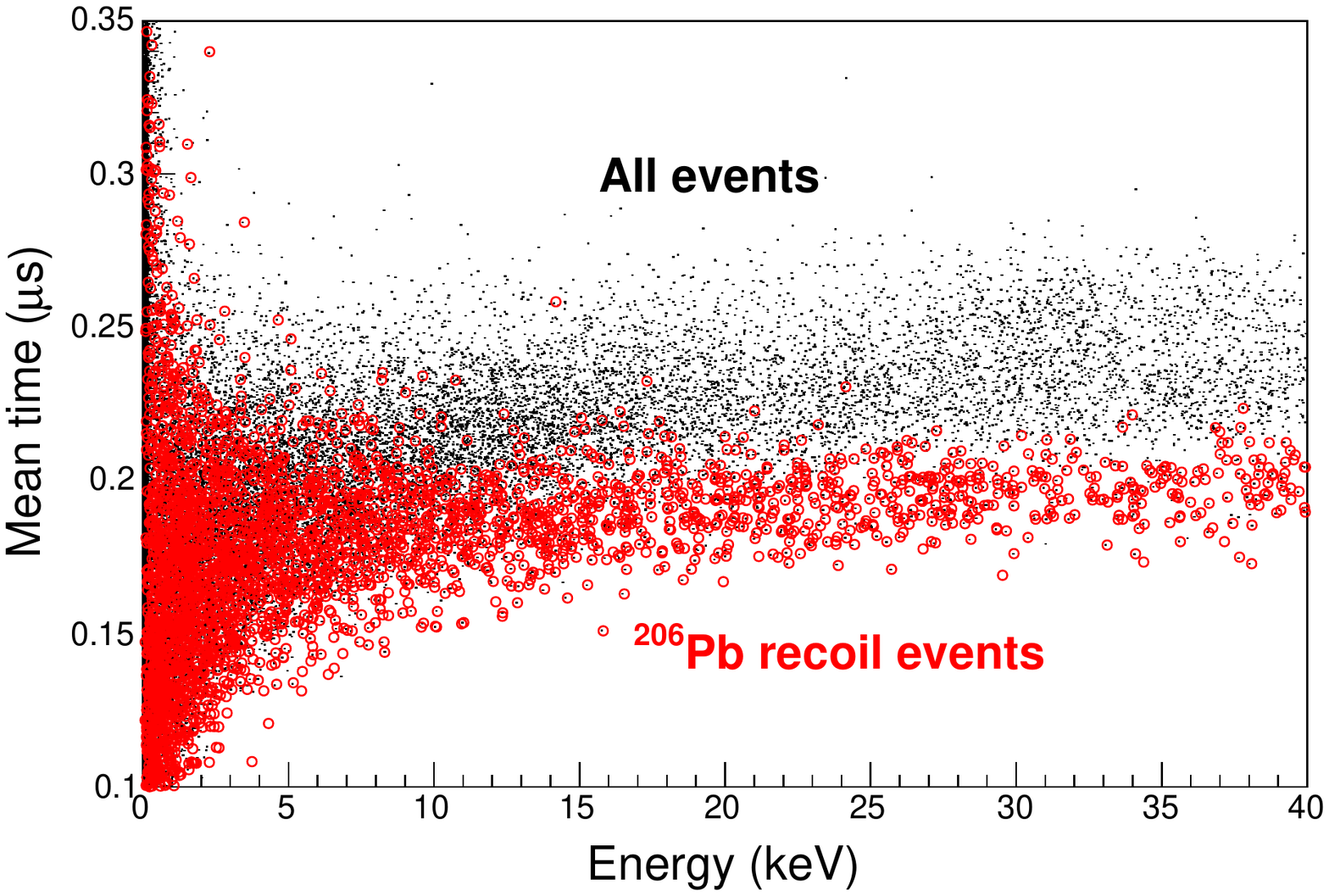}
		\caption{ Mean time distributions as a function of energy for all events (black dot) and the selected \pbsix recoil events from the contaminated crystal B. The surface \pbsix recoil events are selected by requiring coincidence events with escaped $\alpha$ tagging from the clean crystal A (red dots in the Figure~\ref{pb206_nEMT}).} 
\label{pb206_cEMT}
\end{center}
\end{figure}

\begin{figure}[!htb]
\begin{center}
		\includegraphics[width=8cm]{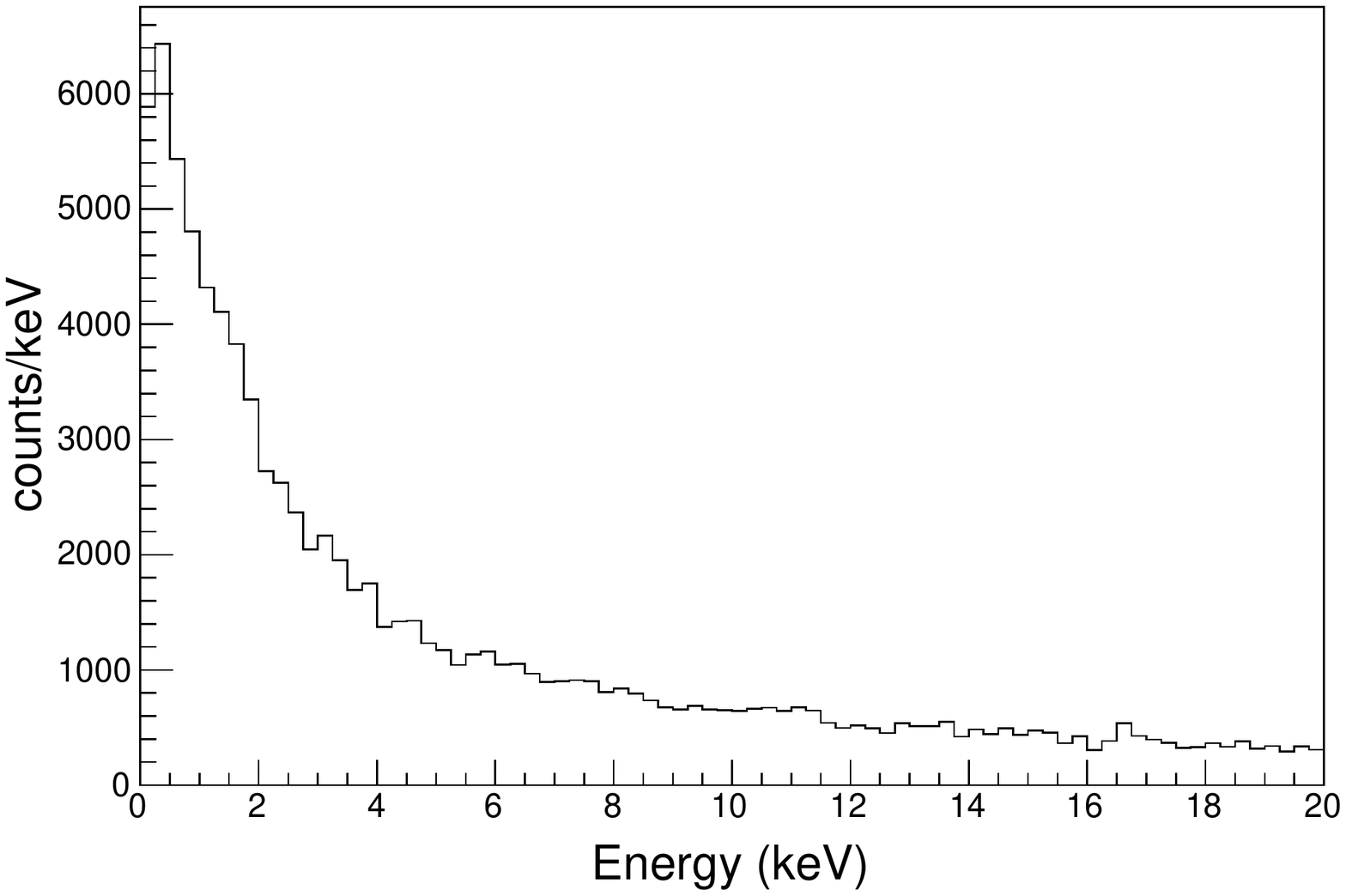}
		\caption{Energy spectrum of the surface \pbsix recoil events.}
\label{pb206_srE}
\end{center}
\end{figure}

When the clean crystal A tagged the escaped $\alpha$ particles, the recoiled \pbsix deposited its kinetic energy in the contaminated crystal B. 
Figure~\ref{pb206_cEMT} shows the mean time distribution as a function of energy in the crystal B for all events and the surface \pbsix recoil events. 
It is clear that the \pbsix recoil events have shorter decay times than typical electron recoil background events. This behavior is similar to the nuclear recoil events in the NaI(Tl) crystal as measured with a neutron source~\cite{kims_psd}. 
The energy spectrum of the selected \pbsix recoil events is shown in Fig.~\ref{pb206_srE}. 
Considering a 103~keV kinetic energy of the \pbsix recoil and quenching factor of iodine nuclear recoil events 5\%~\cite{nai_qf}, one can expect an energy peak around 5~keV. However, the measurement shows a continuous spectrum with increases at lower energy. This was already observed for the CsI(Tl) crystal~\cite{kims_sa}.  It was considered due to the inactive scintillation layer on the crystal surface, but further studies are required for accurate understanding.

\begin{figure*}
\begin{center}
  \begin{tabular}{cc}
  \includegraphics[width=0.5\textwidth]{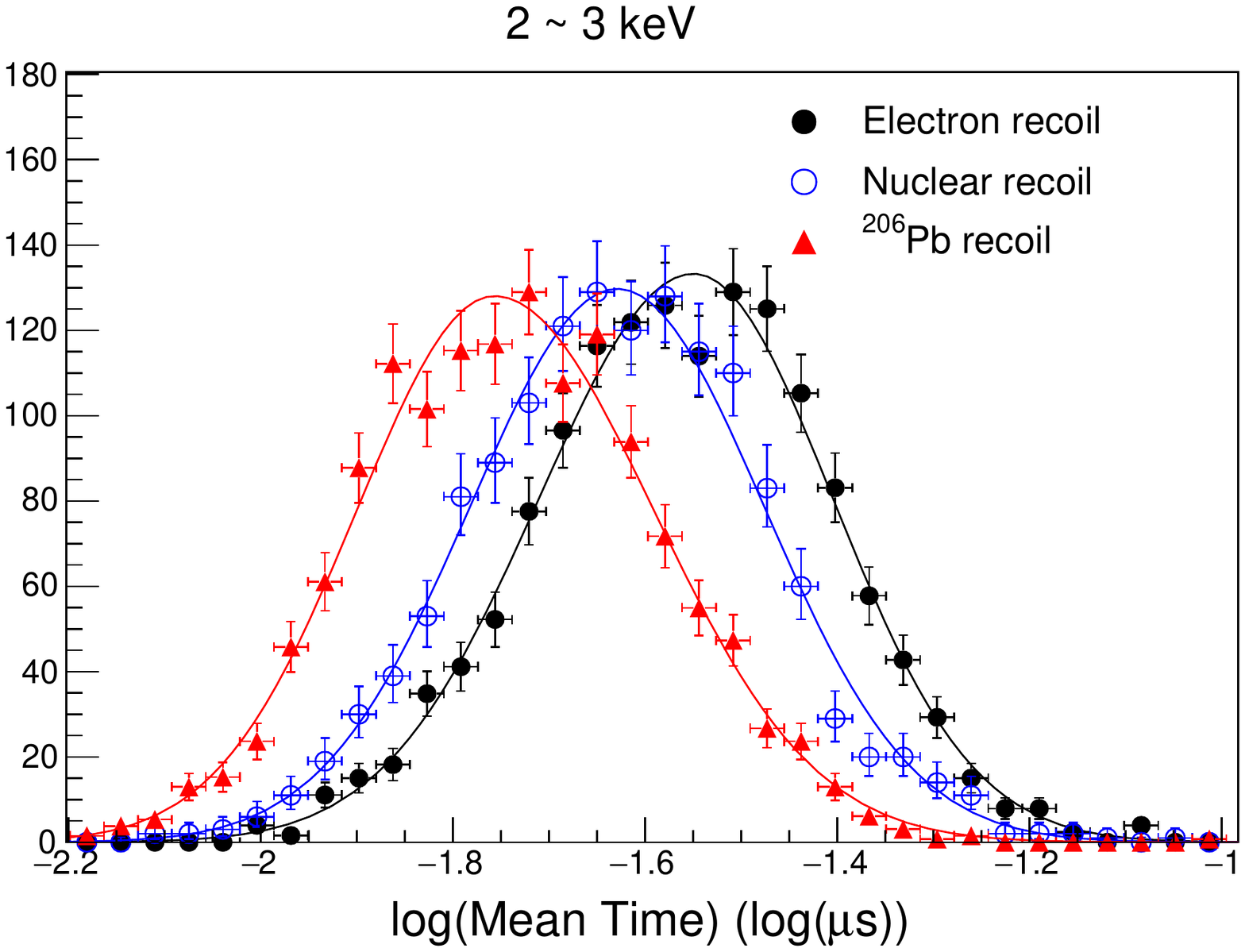} &
   \includegraphics[width=0.5\textwidth]{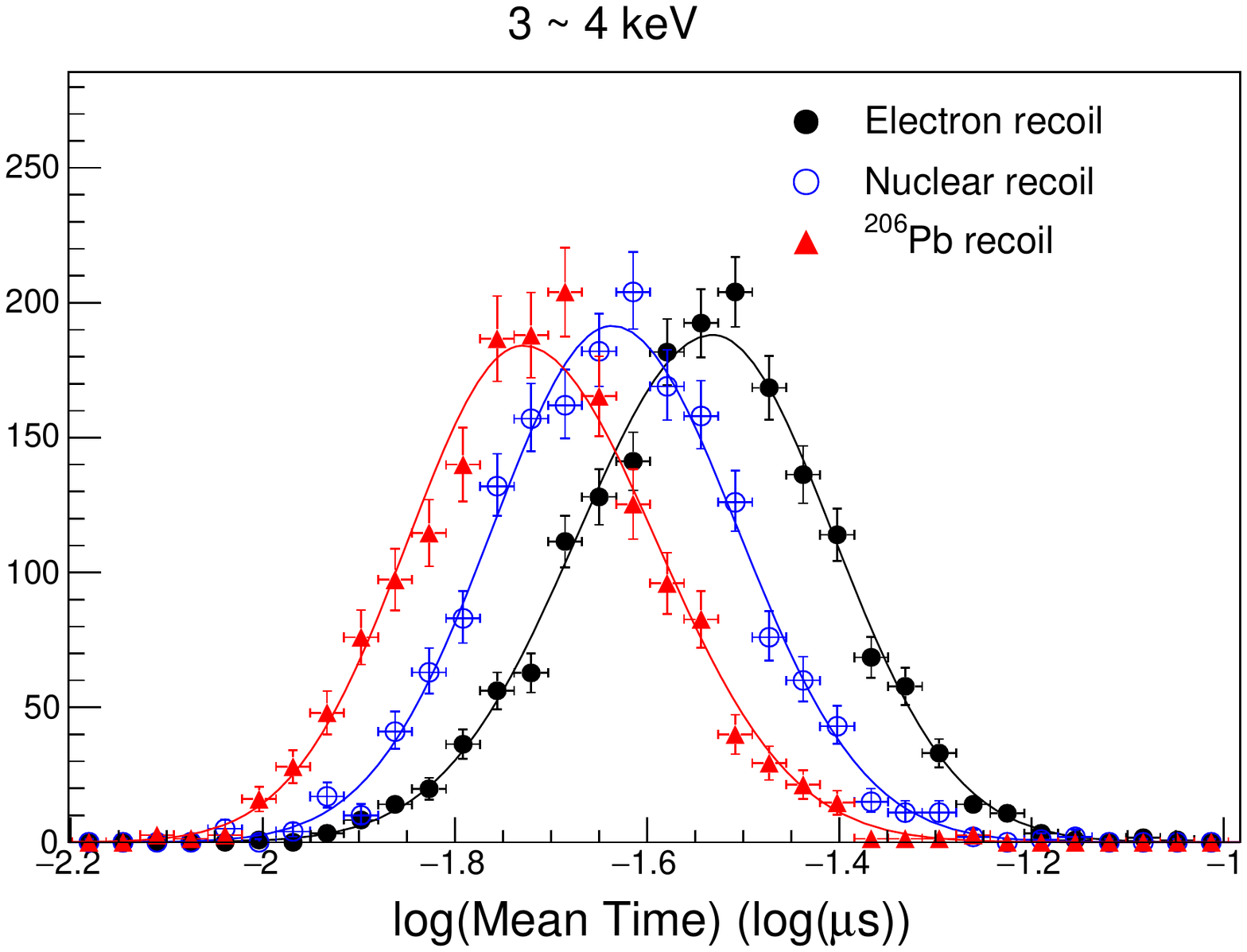} \\
      (a) & (b) \\
   \end{tabular}
		\caption{ Logarithmic mean time distributions for three different types of recoil events (electron recoil, nuclear recoil from neutron calibration, and surface nuclear recoil due to \pbsix recoil on the crystal surface) for 2-3~keV~(a) and 3-4~keV~(b). }
\label{lmt_ernrsr}
\end{center}
\end{figure*}

To quantify the PSD between the \pbsix, nuclear, and electron recoil events, we directly compare the logarithm~(log) of the mean time spectra for 2-3~keV and 3-4~keV visible energy events in Fig.~\ref{lmt_ernrsr}. 
To obtain the nuclear recoil events, the Am-Be neutron source data described in Ref.~\cite{kims_psd} is used. The electron recoil events are collected by Compton scattering events with 662.1~keV $\gamma$-rays from a $^{137}$Cs source. 
In Fig.~\ref{lmt_ernrsr}, differences in the log(mean time) distributions between the \pbsix recoil, the nuclear recoil, and the electron recoil events can be noticed. In particular, clear differences between the \pbsix recoil and typical NaI(Tl) nuclear recoils are evident.

\begin{figure*}
\begin{center}
  \begin{tabular}{cc}
		\includegraphics[width=0.5\textwidth]{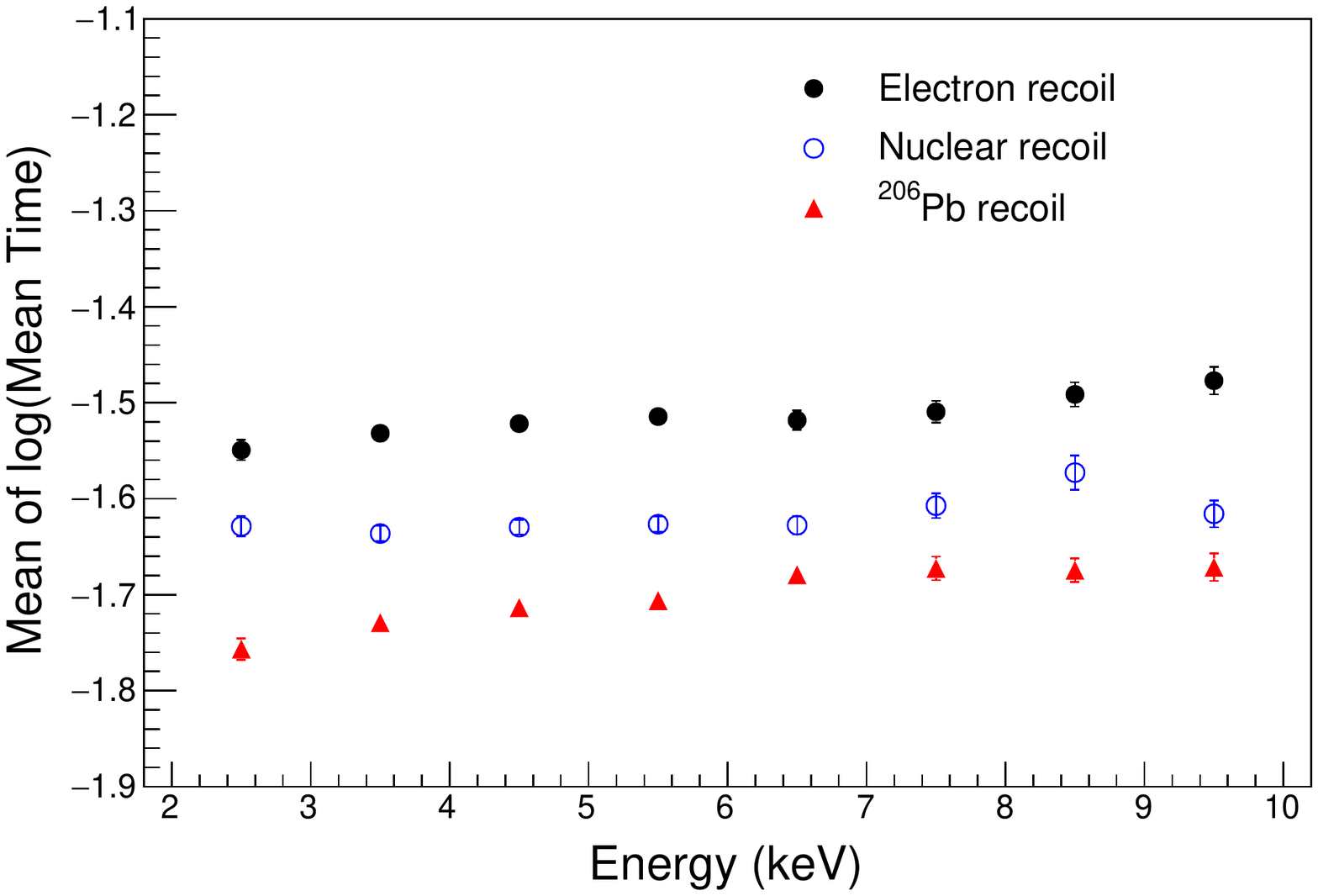} &
		\includegraphics[width=0.5\textwidth]{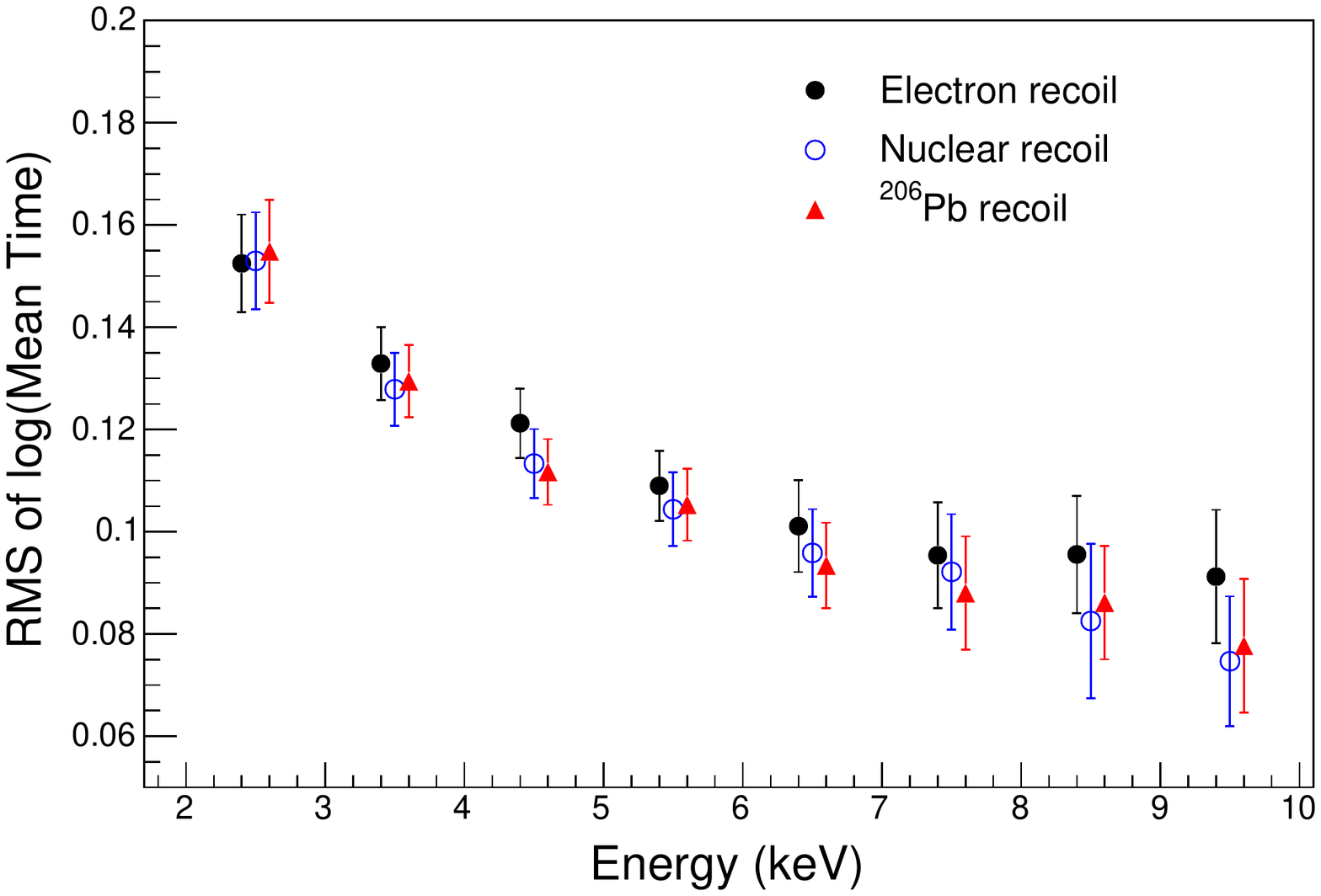}\\
		(a) & (b) \\
  \end{tabular}
		\caption{(a) Mean of the log(mean time) and (b) root-mean-square~(RMS) of the log(mean time) from 2 keV to 10 keV electron, nuclear, and surface \pbsix recoil events are compared. The surface \pbsix events have faster decay time than the nuclear recoil events on average while the RMS are consistent. }
\label{lmt_mean} 
\end{center}
\end{figure*}

		Chi-square fits to the logarithm of the mean time distributions were performed using asymmetric Gaussian functions that are overlaid in Fig.~\ref{lmt_ernrsr}. Good agreement with the data was observed. The same fits for each 1~keV energy bin were carried out for each type of recoil event. 
Figure~\ref{lmt_mean} shows the mean values as well as the root-mean-square~(RMS) of the fits for \pbsix~(surface nuclear recoil), nuclear, and electron recoil events. 
The surface \pbsix recoil events have faster decay times than the electron and nuclear recoil events of sodium and iodine. This suggests that a sufficiently large set of the nuclear recoil events might be extracted statistically from the data even though surface \pbsix recoil events are presented.

\begin{figure}[!htb]
\begin{center}
		\includegraphics[width=8cm]{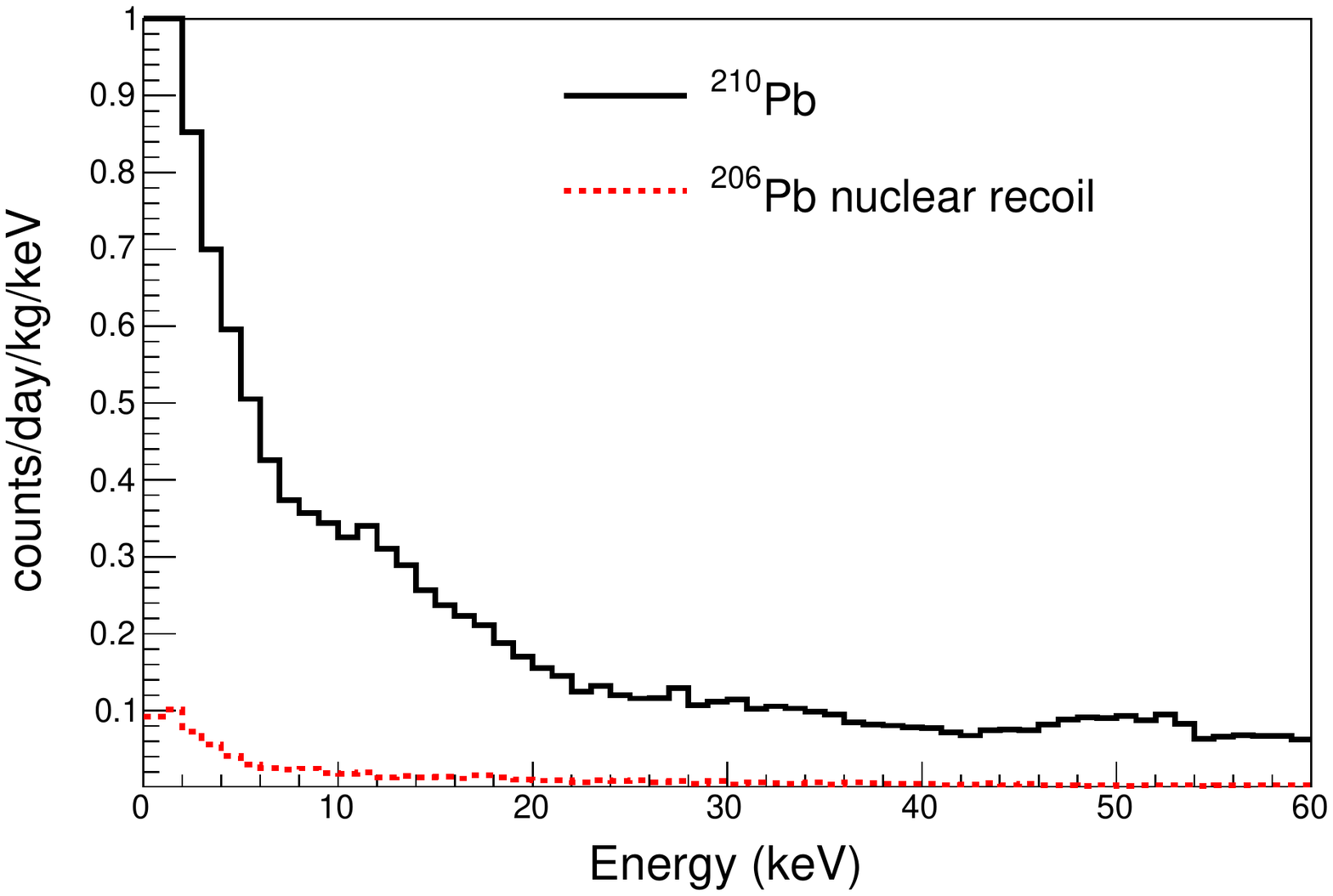}
		\caption{Low-energy spectrum of \pbten and \pbsix recoil events obtained using single-crystal trigger, normalized to the background of NaI-002 crystal in Ref.~\cite{kims_nai}.}
\label{pbspect}
\end{center}
\end{figure}

After three months of measuring \pbsix recoil events, we switched the trigger without the coincidence of the two crystals because the \pbten $\beta$-decays create electron recoil events in a single crystal. We collected data for about a week with the single crystal trigger condition. Using analysis software, we can select the \pbsix events by requiring these events be coincident with the escaped $\alpha$ particles. We found that the rate and energy spectrum of \pbsix recoil events obtained using a single-crystal trigger were consistent with that of earlier coincident-trigger data for energies greater than 1~keV. 
The relative energy spectrum of the \pbten $\beta$ events and the \pbsix recoil events are presented in Fig.~\ref{pbspect}. Plot is normalized to the measured background of a NaI-002 crystal in Ref.~\cite{kims_nai} by multiplying the spectra with the ratio of the surface $\alpha$ rate in Ref.~\cite{kims_nai} to that measured here. 
The measured spectrum and the rate of surface \pbten events are similar to results from simulation-based surface \pbten events~\cite{kims_jeon,anais_bg}.
However, this spectrum contains other background events due to internal contamination and external radioisotopes from the PMTs. 
Detailed simulation-based studies including all of known backgrounds as well as reasonable modeling of the depth profile for the surface \pbten may provide a good understanding of this important background, and effort is ongoing.

\section{Conclusion}
The measurement of surface \pbsix recoil and surface \pbten $\beta$-decay events in the NaI(Tl) crystal were carried out by exposing the NaI(Tl) crystal to a \rntt radioactive source.
The surface \pbsix recoil events have a significantly different mean of the decay time of the scintillating signals than the nuclear recoil of sodium and iodine.
The measured mean time characteristics of the \pbsix recoil events will be used to extract the WIMP-nuclear interaction signals from the COSINE-100 or KIMS-NaI experiments. 

\section*{Acknowledgements}
We thank N. Carlin of University of S\~{a}o Paulo for useful comments. We thank the Korea Hydro and Nuclear Power (KHNP) Company for providing the underground laboratory space at Yangyang. This work was supported by IBS-R016-A1.


\end{document}